\documentclass[runningheads]{llncs}
\usepackage{graphicx}
\usepackage{amsmath}
\usepackage{threeparttable} 
\usepackage{url}
\usepackage{hyperref}

\begin{document}
\title{Breast Ultrasound Tumor Classification Using a Hybrid Multitask CNN-Transformer Network}
\author{Bryar Shareef, Min Xian, Aleksandar Vakanski, Haotian Wang}
\titlerunning{BUS tumor classification using multitask CNN-transformer network}
%

%
\authorrunning{B. Shareef et al.}
%
\institute{University of Idaho}
\institute{Department of Computer Science, University of Idaho, Idaho Falls, Idaho 83402, USA}

\maketitle              
\begin{abstract}
Capturing global contextual information plays a critical role in breast ultrasound (BUS) image classification. Although convolutional neural networks (CNNs) have demonstrated reliable performance in tumor classification, they have inherent limitations for modeling global and long-range dependencies due to the localized nature of convolution operations. Vision Transformers have an improved capability of capturing global contextual information but may distort the local image patterns due to the tokenization operations. 
In this study, we proposed a hybrid multitask deep neural network called Hybrid-MT-ESTAN, designed to perform BUS tumor classification and segmentation using a hybrid architecture composed of CNNs and Swin Transformer components. The proposed approach was compared to nine BUS classification methods and evaluated using seven quantitative metrics on a dataset of 3,320 BUS images. The results indicate that Hybrid-MT-ESTAN achieved the highest accuracy, sensitivity, and F1 score of 82.7\%, 86.4\%, and 86.0\%, respectively.

\end{abstract}
\section{Introduction}
Breast cancer is the leading cause of cancer-related fatalities among women. Currently, it holds the highest incidence rate of cancer among women in the U.S., and in 2022 it accounted for 31\% of all newly diagnosed cancer cases \cite{networks}. Due to the high incidence rate, early breast cancer detection is essential for reducing mortality rates and expanding treatment options. BUS imaging is an effective screening option because it is cost-effective, nonradioactive, and noninvasive. However, BUS image analysis is also challenging due to the large variations in tumor shape and appearance, speckle noise, low contrast, weak boundaries, and occurrence of artifacts.

In the past decade, deep learning-based approaches achieved remarkable advancements in BUS tumor classification \cite{zhuang2021breast,shareef2023Benchmark}. The progress has been driven by the capability of CNN-based models to learn hierarchies of structured image representations as semantics. To extract deep context features, CNNs apply a series of convolutional and downsampling layers, frequently organized into blocks with residual connections. Nevertheless, one disadvantage of such architectural choice is that the feature representations in the deeper layers become increasingly abstract, leading to a loss of spatial and contextual information. The intrinsic locality of convolutional operations hinders the ability of CNNs to model long-range dependencies while preserving spatial information in images effectively.

Vision Transformer (ViT) \cite{dosovitskiy2020image} and its variants recently demonstrated superior performance in image classification tasks. These models convert input images into smaller patches and utilize the self-attention mechanism to model the relationships between the patches. Self-attention enables ViTs to capture long-range dependencies and model complex relationships between different regions of the image. However, the effectiveness of ViT-based approaches heavily relies on access to large datasets for learning meaningful representations of input images. This is primarily because the architectural design of ViTs does not rely on the same inductive biases in feature extraction which allow CNNs to learn spatially invariant features.  

Accordingly, numerous prior studies introduced modifications to the original ViT network specifically designed for BUS image classification \cite{gheflati2022vision,ayana2022buvitnet,hassanien2022predicting}. In addition, several works proposed network architectures that combined Transformers and CNNs \cite{mo2022hover,qu2022vgg,BTSST2023}. For instance, Mo et al. \cite{mo2022hover} proposed a hybrid CNN-Transformer incorporating BUS anatomical priors. Qu et al. \cite{qu2022vgg} employed squeeze and excitation blocks to enhance the feature extraction capacity in a hybrid CNN-based VGG16 network and ViT. Similarly, Iqbal et al. \cite{BTSST2023} designed two hybrid CNN-Transformer networks intended either for classification or segmentation of multi-modal breast cancer images. Despite the promising results of such hybrid approaches, effectively capturing the local patterns and global long-range dependencies in BUS images remains challenging \cite{BTSST2023,dosovitskiy2020image,tang2023transformer}.  

Multitask learning leverages shared information across related tasks by jointly training the model. It constrains models to learn representations that are relevant to all tasks rather than learning task-specific details. Moreover, multitask learning acts as a regularizer by introducing inductive bias and prevents overfitting \cite{Sebastian2017} (particularly with ViTs), and with that, can mitigate the challenges posed by small BUS dataset sizes. In \cite{shareef2023Benchmark}, the authors demonstrated that multitask learning outperforms single-task learning approaches for BUS classification.

In this study, we introduce a hybrid multitask approach, Hybrid-MT-ESTAN, which encompasses tumor classification as a primary task and tumor segmentation as a secondary task. Hybrid-MT-ESTAN combines the advantages of CNNs and Transformers in a framework incorporating anatomical tissue information in BUS images. Specifically, we designed a novel attention block named Anatomy-Aware Attention (AAA), which modifies the attention block of Swin Transformer by considering the breast anatomy. The anatomy of the human breast is categorized into four primary layers: the skin, premammary (subcutaneous fat), mammary, and retromammary layers, where each layer has a distinct texture and generates different echo patterns. The primary layers in BUS images are arranged in a vertical stack, with similar echo patterns appearing horizontally across the images. The kernels in the introduced AAA attention blocks are organized in rows and columns to capture the anatomical structure of the breast tissue. In the published literature, the closest approach to ours is the work by Iqbal et al.  \cite{BTSST2023}, in which the authors used hybrid single-task CNN-Transformer networks for either classification or segmentation of BUS images. Conversely, Hybrid-MT-ESTAN employs a multitask approach and introduces novel architectural design. The main contributions of this work are summarized as:
\begin{itemize}
    \item[•] The proposed architecture effectively integrates the advantages of CNNs for extracting hierarchical and local patterns in BUS images and Swin Transformers for leveraging long-range dependencies.
    \item[•] The designed Anatomy-Aware Attention (AAA) block improves the learning of contextual information based on the anatomy of the breast.
    \item[•] The multitask learning approach leverages the shared representations across the classification and segmentation tasks to improve the model performance.
\end{itemize}

\vspace*{-6mm}  
\begin{figure}[h]
\includegraphics[width=0.94\linewidth]{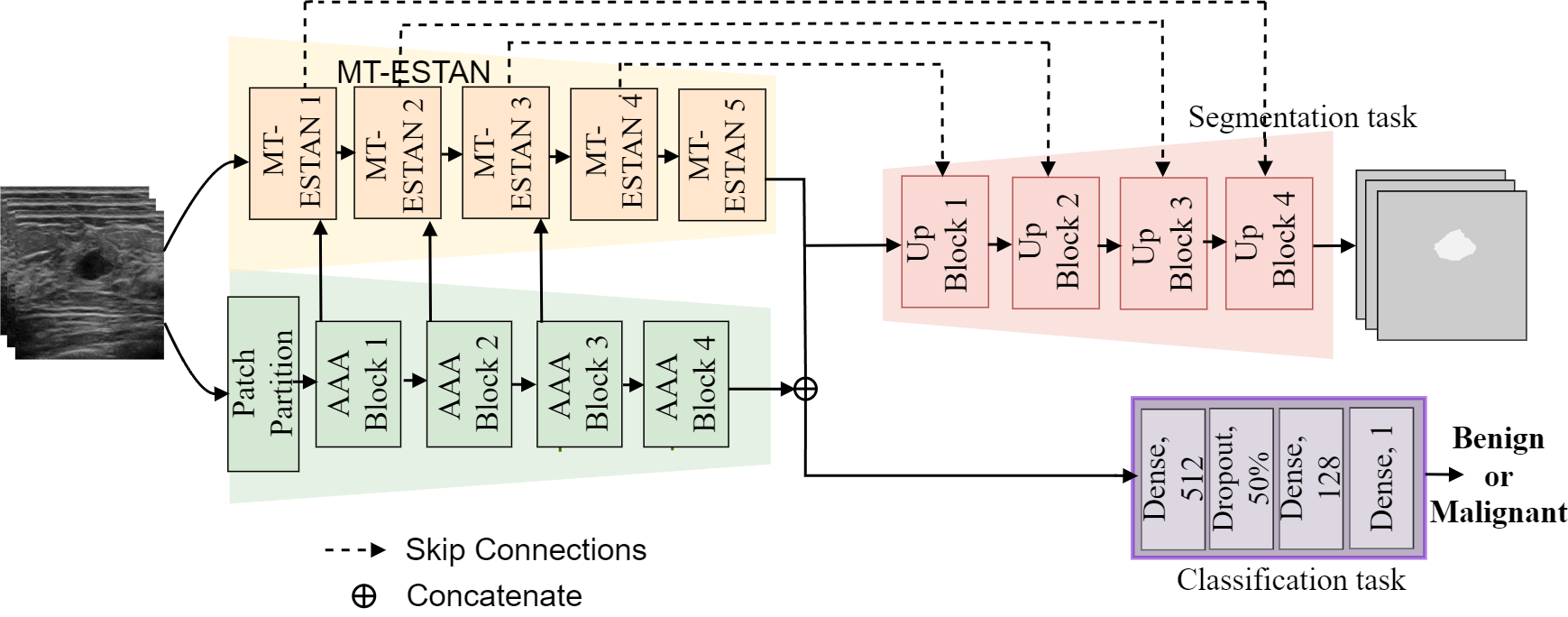}

\caption{Hybrid-MT-ESTAN consists of MT-ESTAN and AAA encoders, a segmentation branch, and a classification branch.} \label{fig2}
\vspace*{-3mm}
\end{figure}
\vspace*{-6mm}
\section{Proposed Method}
\subsection{Hybrid-MT-ESTAN}
\begin{figure}[h]
\centering
\includegraphics[width=0.25\linewidth]{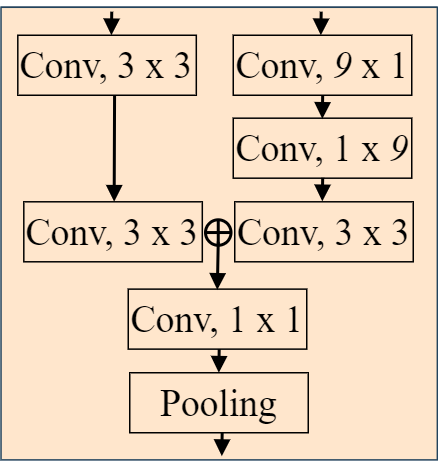}
\caption{ MT-ESTAN blocks include parallel convolutional branches with different kernel size, followed by 1x1 convolution and a pooling layer.} \label{fig4}
\end{figure}

The architecture of Hybrid-MT-ESTAN is shown in Fig. \ref{fig2}, and consists of: (1) the MT-ESTAN encoder \cite{shareef2023Benchmark}, and a Swin Transformer-based encoder with Anatomy-Aware Attention (AAA) blocks, (2) a decoder branch for the segmentation task, and (3) a branch with fully-connected layers for the classification task. MT-ESTAN \cite{shareef2023Benchmark} is a CNN-based multitask learning network that simultaneously performs BUS classification and segmentation. The encoder sub-network of MT-ESTAN is ESTAN \cite{shareef2020estan}, which employs row-column-wise kernels to learn and fuse context information in BUS images at different context scales (see Fig. \ref{fig4}). Specifically, each MT-ESTAN block is composed of two parallel branches consisting of four square convolutional kernels and two consecutive row-column-wise kernels. These specialized convolutional kernels effectively extract contextual information of small tumors in BUS images. Refer to \cite{shareef2020estan}, \cite{shareef2020stan}, and \cite{shareef2023Benchmark} for the implementation details of ESTAN and MT-ESTAN. The source codes of these works are available at \href{http://busbench.midalab.net}{http://busbench.midalab.net}. 
\vspace*{-5mm}
\vspace*{-11mm}

\subsection{Anatomy-Aware Attention (AAA) Block}
\begin{figure}[h]
\includegraphics[width=0.94\linewidth]{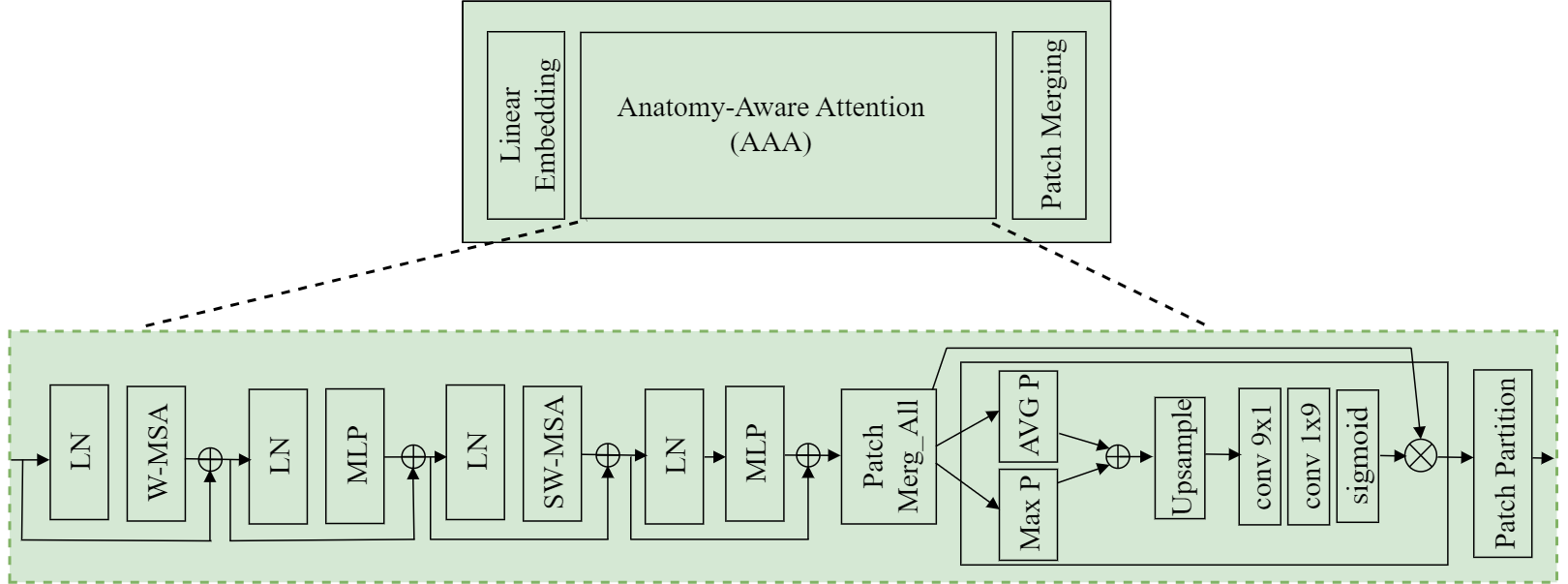}
\caption{ Anatomy-Aware Attention (AAA) block.} \label{fig3}
\vspace*{-2mm}
\end{figure}

Swin Transformer \cite{liu2021swin} is a hierarchical transformer-based approach that uses shifted windows to model global context information. Swin Transformer partitions an input image into non-overlapping patches of size $4\times 4$, where each patch is treated as a "token". A linear layer receives the patches and projects them into an arbitrary dimension. Each Swin Transformer block consists of a LayerNorm layer (LN) layer, a multi-head self-attention module (MSA), and a multi-layer perceptron (MLP) with GELU activation. To model long-range dependencies, the original Swin Transformer relies on shifted windows, where the window-based multi-head self-attention (W-MSA) and shifted window-based multi-head self-attention (SW-MSA) modules are employed in each consecutive Swin block. The Swin block is formulated as follows.
\\
\begin{align}
 \hat{f}^l=\textsc{W-MSA}(\textsc{LN}({f}^{l-1})) + {f}^{l-1} \\ {f}^{l} =\textsc{MLP}(\textsc{LN}(\hat{f}^{l})) + \hat{f}^{l}\\
\hat{f}^{l+1} = \textsc{SW-MSA}(\textsc{LN}({f}^{l})) + {f}^{l} \\ 
{f}^{l+1} =\textsc{MLP}(\textsc{LN}(\hat{f}^{l+1})) + \hat{f}^{l+1} 
\end{align}
where $f^{l}$ and $\hat{f}^{l}$ are the output features of the MLP module and the (S)W-MSA module for block $l$, respectively;
in the proposed Anatomy-Aware Attention (AAA) block, we redesigned the Swin blocks to enhance their ability to model both global and local features by adding an attention block based on the breast anatomy (see Fig. \ref{fig3}). The additional layers are defined by 
\\
\begin{align}
{y}^{i} =M({f}^{l+1})\\
{B}^{i} =U(\textsc{MAX-P}({y}^{i})+\textsc{AVG-P}({y}^{i}))\\
{O}^{i} ={y}^{i} \cdot (\sigma(A(B)))
\end{align}

Concretely, we first reconstruct the $i$-th feature map ($y^{i}$) by merging ($M$) all patches, and afterward, we applied average pooling (\textsc{AVG-P}) and max pooling (\textsc{MAX-P}) layers with size (2, 2). The outputs of (\textsc{AVG-P}) and (\textsc{MAX-P}) layers are concatenated and  \textsc{up-sampled ($U$)} with size (2, 2) and stride (2, 2). \textsc{Row-column-wise} kernels ($A$) with size (9 , 1) and (1 , 9) are then employed to adapt to the anatomy of the breast, and finally a sigmoid function $(\sigma)$ is applied to the output of (A) multiplied by the input feature map ($y^{i}$).
\vspace*{-4mm}
\vspace*{-3mm}
\subsection{Segmentation and Classification Branches/Tasks}
\label{ssec:subhead}
The segmentation branch in Fig. \ref{fig2} outputs dense mask predictions of BUS tumors. It consists of four Up Blocks, each with three convolutional layers and one upsampling layer (with size (2, 2) and stride (2, 2)). The settings of the convolutional layers are adopted from \cite{shareef2023Benchmark}. In addition, the blocks receive four skip connections from the MT-ESTAN encoder, i.e., there is a skip connection from each MT-ESTAN block 1 to 4. The classification branch consists of three dense layers, a dropout layer (50\%), and the final dense layer that predicts the tumor class into benign or malignant.

\label{ssec:subhead}
\subsection{Loss Function}
We applied a multitask loss function ($L_{mt}$) that aggregates two terms: a focal loss $L_{Focal}$  for the classification task and dice loss $L_{Dice}$ for the segmentation task. Therefore, the composite loss  function is $L_{mt}= w_1\cdot L_{Focal}+ L_{Dice}$, where the weight coefficient $w_1$ is set to apply greater importance to the classification task as the primary task. Since in medical image diagnosis achieving high sensitivity places emphasis on the detection of malignant lesions, we employed the focal loss for the classification task to trade off between sensitivity and specificity. Because malignant tumors are more challenging to detect due to greater differences in margin, shape, and appearance in BUS images, focal loss forces the model to focus more on difficult predictions. Specifically, focal loss adds a factor $(1-p_i)^\gamma$ to the cross-entropy loss where $\gamma$ is a focusing parameter, resulting in $L_{Focal}= -1/N \sum_{i=1}^N[(\alpha \cdot t_i \cdot (1-p_i )^\gamma \cdot log(p_i) + (1 -\alpha) \cdot p_i \cdot log(1-p_i)]$. In the formulation,  $\alpha$ is a weighting coefficient, $N$ denotes the number of image samples, $t_i$ is the target label of the $i^{th}$ training sample, and $p_i$ denotes the prediction. The segmentation loss is calculated using the commonly-employed Dice loss ($L_{Dice}$) function.

\section{Experimental Results}
\label{sec:pagestyle}

 \subsection{Datasets}We evaluated the performance of Hybrid-MT-ESTAN using four public datasets, HMSS \cite{hmss}, BUSI \cite{ref_article1}, BUSIS \cite{zhang2022busis}, and Dataset B \cite{yap2017automated}. We combined all four datasets to build a large and diverse dataset with a total of 3,320 B-mode BUS images, of which 1,664 contain benign tumors and 1,656 have malignant tumors. Table 1 shows the detailed information for each dataset. HMSS dataset does not provide the segmentation ground-truth masks, and for this study we arranged with a group of experienced radiologists to prepare the masks for HMSS. Refer to the original publications of the datasets for more details.
 \vspace*{-3mm}
  
\begin{table}

\small
\begin{center}
\caption{Breast ultrasound (BUS) datasets. 'b' denotes benign tumor and 'm' is malignant tumor.}
\vspace*{-2 mm}
\begin{tabular}{|l|l|l|l|}
\hline
BUS dataset& No. of images &Distribution &Source                           
\\ \hline
 HMSS   & 1,948 & b:812, m:1136 & Netherlands\\
 BUSI &  647  & b:437, m:210 & Egypt\\
 BUSIS & 562 & b:306, m:256 &  China\\
 Dataset B    &163 & b:109, m:54 &  Spain\\
 \hline
 Total & 3,320 & b: 1,664, m: 1,656&\\ 
 
 \hline
\end{tabular}
\end{center}
\vspace*{-7mm}
\end{table}
\vspace*{-7mm}
\subsection{Evaluation Metrics} For performance evaluation of the classification task, we used the following metrics: accuracy (Acc), sensitivity (Sens), specificity (Spec), F1 score, Area Under the Curve of Receiver Operating Characteristic (AUC), false positive rate (FPR), and false negative rate (FNR). To evaluate the segmentation performance, we used dice similarity coefficient (DSC) and Jaccard index (JI).

\subsection{Implementation Details} 
The proposed approach was implemented with Keras and TensorFlow libraries. All experiments were performed on a machine with NVIDIA Quadro RTX 8000 GPUs and two Intel Xeon Silver 4210R CPUs (2.40GHz) with 512 GB of RAM.  All BUS images in the dataset were zero-padded and reshaped to form square images. To avoid data leakage and bias, we selected  the train, test, and validation sets based on the cases, i.e., the images from one case (patient) were assigned to only one of the training, validation, and test sets. Furthermore, we employed horizontal flip, height shift (20\%), width shift (20\%), and rotation (20 degrees) for data augmentation. The proposed approach utilizes the building blocks of ResNet50 and Swin-Transformer-V2, pretrained on ImageNet dataset. Namely, MT-ESTAN uses pretrained ResNet50 as a base model for the five encoder blocks (the implementation details of MT-ESTAN can be found in \cite{shareef2023Benchmark}). The encoder with AAA blocks uses the SwinTransformer\_V2\_Base\_256 pretrained model as a backbone. For the composite loss function, we adopted a weight coefficient $w_1=3$, and in the focal loss $\alpha=0.5$ and $\gamma=2$. For model training we utilized Adam optimizer with a learning rate of $10^{-5}$ and mini batch size of 4 images.
\vspace*{-2mm}

\subsection{Performance Evaluation and Comparative Analysis}
\label{sssec:subhead}
\vspace*{-1mm}

We compared the performance of Hybrid-MT-ESTAN for BUS classification to nine deep learning approaches commonly used for medical image analysis. The compared models include CNN-based, ViT-based, and hybrid approaches. CNN-based networks are SHA-MTL \cite{zhang2021sha}, MobileNet \cite{howard2017mobilenets}, DenseNet121 \cite{huang2017densely}, and EMT-Net \cite{shi2022emt}. ViT-based approaches include the original ViT \cite{dosovitskiy2020image}, Chowdery \cite{ref_article2}, and Swin Transformer \cite{liu2021swin}. VGGA-ViT \cite{qu2022vgg} is a hybrid CNN-Transformer network. The values of the performance metrics are shown in Table \ref{table2}, indicating that the proposed Hybrid-MT-ESTAN outperformed all nine approaches by achieving the best accuracy, sensitivity, F1 score, and AUC with 82.8\%, 86.4\%, 86.0\%, and 82.8\%, respectively. Although SHA-MTL \cite{zhang2021sha} obtained the highest specificity of 90.8\% and FPR of 9.2\%, the trade-off between sensitivity and specificity should be taken into consideration, as that approach had sensitivity of 48.1\%. The preferred trade-off in medical image analysis typically is high sensitivity without significant degradation in specificity.\\
We evaluated the segmentation performance of Hybrid MT-ESTAN and compared the results to five multitask approaches, including SHA-MTL \cite{zhang2021sha}, EMT-Net \cite{shi2022emt}, Chowdery \cite{ref_article2}, MT-ESTAN \cite{shareef2023Benchmark}, and VGGA-ViT \cite{qu2022vgg}. As shown in Table \ref{table2},the proposed Hybrid MT-ESTAN achieved the highest performance and increased DSC and JI by 5.9\% and 6.4\%, respectively compared to MT-ESTAN. Note that results of single-task models in Table \ref{table2} are not provided.

\begin{table}[t]
\small
\caption{Performance metrics of the compared methods for BUS image classification and segmentation.}
\vspace*{-3 mm}
\label{table2}

\begin{center}
\begin{tabular}{|l|l|l|l|l|l|l|l||l|l|}

\hline
 & \multicolumn{7}{l|}{Classification} & \multicolumn{2}{l|}{Segmentation} \\
\hline

Methods & Acc$\uparrow$ & Sens.$\uparrow$ & Spec.$\uparrow$ & F1$\uparrow$ & Auc$\uparrow$  & FNR$\downarrow$ & FPR$\downarrow$ & DSC$\uparrow$& {  JI$\uparrow$} \\
\hline

SHA-MTL \cite{zhang2021sha} &69.6	&48.1	&\textbf{90.8}	&0.58	&69.5	&51.9	&\textbf{9.2} & 72.2	&60.7

\\
MobileNet \cite{howard2017mobilenets} & 71.0	 &82.0	 &61.0	 &0.74	 &71.5	 &18.0	 &39.0 & { -}	& { -}

\\
VGGA-ViT \cite{qu2022vgg} &73.6	&61.8	&79.8	&0.61	&70.8	&38.2	&20.2 &74.9	& 64.9

 \\
DenseNet121 \cite{huang2017densely} & 73.0	& 74.0	& 71.0	& 0.73	& 72.5	& 26.0	& 29.0 &{ -}& { -}
\\

EMT-Net \cite{shi2022emt} & 74.1 &79.4	& 69.1	& 0.75	& 74.3	& 20.6	& 30.9 &76.7
& 67.0
 \\
 
ViT \cite{dosovitskiy2020image} & 72.1	&74.1	&69.3	&0.73	&71.7	&25.9	&30.7 &{ -}& { -}
\\

Chowdery \cite{ref_article2} &77.4	&77.3&	77.3	&0.77	&77.3&	22.7&	22.7 &77.0	&67.9

\\
Swin Transformer & 77.4	&72.6	&82.5	&0.74	&77.6	&27.4	&17.5  &{ -}& { -} 
\\
MT-ESTAN  &78.6	&83.7	&72.6	&0.83	&78.2	&16.3	&27.4 &{78.2}	&{69.3}
\\
Ours   &\textbf{82.8}	&\textbf{86.4}	&79.2	&\textbf{0.86}	&\textbf{82.8}	&\textbf{13.6}	&20.8 &\textbf{84.1} &	\textbf{75.7}
\\ 
\hline  

\end{tabular}
     \begin{tablenotes}
       \item [1] Note: A dash '-' in the Segmentation column indicates that the model uses single-task learning. 
       
     \end{tablenotes}
\end{center}
\vspace*{-6mm}
\vspace*{0.1mm}
\end{table}

\begin{table}[t]
\small
\caption{Ablation study for evaluating the components of Hybrid-MT-ESTAN.}
\vspace*{-10mm}
\begin{center}
\label{tabl3}
\begin{tabular}{|l|l|l|l|l|l|l|l||l|l|}
\hline
 & \multicolumn{7}{l|}{Classification} & \multicolumn{2}{l|}{Segmentation} \\
\hline
Methods & Acc$\uparrow$ & Sens.$\uparrow$ & Spec.$\uparrow$ & F1$\uparrow$ & Auc$\uparrow$  & FNR$\downarrow$ & FPR$\downarrow$& DSC$\uparrow$& {  JI$\uparrow$} \\
\hline
MT-ESTAN \cite{ref_article2} &78.6	&83.7	&72.6	&0.83	&78.2	&16.3	&27.4 &78.2	&69.3
\\
Swin Trans. & 77.4	&72.6	&\textbf{82.5}	&0.74	&77.6	&27.4	&17.5 &{ -} & { -}	
\\ 
MT-ESTAN + Swin Trans. &80.3	&84.2	&76.3	&0.83	&80.2	&15.8	&23.7 &82.3	&73.6

 \\
 Ours  &\textbf{82.8}	&\textbf{86.4}	&79.2	&\textbf{0.86}	&\textbf{82.8}	&\textbf{13.6}	&20.8 &\textbf{84.1} &\textbf{75.7}
\\ 

\hline  
\end{tabular}    
\end{center}
\vspace*{-7mm}
\end{table}

\subsection{Effectiveness of the Anatomy-Aware Attention (AAA) Block}
To verify the effectiveness of the Anatomy-Aware Attention (AAA) block, we conducted an ablation study that quantified the impact of the different components in Hybrid-MT-ESTAN on the classification and segmentation performance. Table \ref{tabl3} presents the values of the performance metrics for MT-ESTAN (pure CNN-based approach), Swin Transformer (pure Transformer network), a hybrid architecture of MT-ESTAN and Swin Transformer, and our proposed Hybrid-MT-ESTAN with AAA block. According to the results in Table \ref{tabl3}, MT-ESTAN achieved better sensitivity and F1 score than Swin Transformer, with 83.7\% and 83\%, respectively. The hybrid architectures of MT-ESTAN with Swin Transformer improved the classification performance and has higher accuracy, sensitivity, F1 score, and AUC with 80.3\%, 84.2\%, 83\%, and 80.2\%, compared to MT-ESTAN and Swin Transformer individually. The proposed approach, Hybrid-MT-ESTAN with AAA block, further improved accuracy, sensitivity, F1 score, and AUC by 2.5\%, 2.2\%, 3\%, and 2.6\%, respectively, relative to the hybrid model without the AAA block. 

To evaluate the segmentation performance, we compared the proposed approach with and without the AAA block and Swin Transformer. As shown in Table \ref{tabl3}, MT-ESTAN combined with Swin Transformer improved DSC and JI by 4.1\% and 4.3\%, respectively compared to MT-ESTAN. Employing the proposed AAA block further improved DSC and JI  by 1.8\% and 2.1\%, respectively.   

\section{Conclusion}
\label{sec:conclusion}
In this paper, we introduced the Hybrid-MT-ESTAN, a multitask learning approach for BUS image analysis that alleviates the lack of global contextual information in the low-level layers of CNN-based approaches. Hybrid-MT-ESTAN concurrently performs BUS tumor classification and segmentation, with a hybrid architecture that employs CNN-based and Swin Transformer layers. The proposed approach exploits multi-scale local patterns and global long-range dependencies provided by MT-ESTA  and AAA Transformer blocks for learning feature representations, resulting in improved generalization. Experimental validation demonstrated significant performance improvement by Hybrid-MT-ESTAN in comparison to current state-of-the-art models for BUS classification.   
\vspace*{-16mm}
\section*{}
\textbf{Acknowledgement.} Research reported in this publication was supported by the National Institute Of General Medical Sciences of the National Institutes of Health under Award Number P20GM104420. The content is solely the responsibility of the authors and does not necessarily represent the official views of the National Institutes of Health. 

%
%
%
%

\end{document}